\documentclass[12pt]{article}
\usepackage{amsmath,amssymb,mathrsfs,cite,color}
\usepackage{graphicx}
\usepackage{hyperref}
\usepackage{comment}
\usepackage{float}
\usepackage{here}
\newcommand{\tr}{\operatorname{tr}}
\newcommand{\Tr}{\operatorname{Tr}}

\newcommand{\dif}{d} %{\mathrm{d}}
\newcommand{\der}[2]{\frac{\dif #1}{\dif #2}}
\newcommand{\pder}[2]{\frac{\partial #1}{\partial #2}}
\newcommand{\iDelta}{\mathit{\Delta}}

\allowdisplaybreaks
\numberwithin{equation}{section}

\begin{document}
\begin{titlepage}
\begin{flushright}
% Preprint # %%%%%%%%%%%%%%%%%%%%%
  UTHEP-809, KEK-TH-2747
%\today
\end{flushright}
%%\vspace{12mm}
%%\vspace{8mm}
\vspace{2mm}
\begin{center}
{\Large \bf
% Title %%%%%%%%%%%%%%%%%%%%%%%%
  Quantum Monte Carlo calculations in the nuclear shell model
  by the complex Langevin method
}
\end{center}
\vspace{2mm}
%%\vspace{7mm}
\begin{center}
% Author %%%%%%%%%%%%%%%%%%%%%%%
Yuhma Asano$^{a,b,}$\footnote{
%E-mail:
asano@het.ph.tsukuba.ac.jp
},
Yuta Ito$^{c,}$\footnote{
%E-mail:
y-itou@tokuyama.ac.jp
},
Jun Nishimura$^{d,e,}$\footnote{
%E-mail:
jnishi@post.kek.jp
},
and
Noritaka Shimizu$^{f,}$\footnote{
%E-mail:
shimizu@nucl.ph.tsukuba.ac.jp
}

\par \vspace{7mm}
% Affiliation %%%%%%%%%%%%%%%%%%%%%
$^a${\it
Institute of Pure and Applied Sciences, University of Tsukuba,\\
1-1-1 Tennodai, Tsukuba, Ibaraki 305-8571, Japan
}\\

$^b${\it
Tomonaga Center for the History of the Universe, University of Tsukuba,\\
1-1-1 Tennodai, Tsukuba, Ibaraki 305-8571, Japan
}\\

$^c${\it
National Institute of Technology, Tokuyama College,\\
Gakuendai, Shunan, Yamaguchi 745-8585, Japan
}\\

$^d${\it
KEK Theory Center, High Energy Accelerator Research Organization,\\
1-1 Oho, Tsukuba, Ibaraki 305-0801, Japan
}\\

$^e${\it Graduate Institute for Advanced Studies, SOKENDAI,\\
  1-1 Oho, Tsukuba, Ibaraki 305-0801, Japan
}\\

$^f${\it
  Center for Computational Sciences, University of Tsukuba,\\
  1-1-1 Tennodai, Tsukuba, Ibaraki 305-8571, Japan
}\\

\end{center}
\vspace{2mm}
%%\vspace{7mm}
\begin{abstract}\noindent%%%%%%%%%%%%%%%
  The nuclear shell model
  is known to describe the properties of various nuclei
  extremely well.
  However,
  the auxiliary-field quantum Monte Carlo calculations cannot be applied to it
  with general interactions due to the sign problem.
  The model
  has therefore been investigated primarily by variational methods,
  where the accuracy of the results depends crucially on the ansatz for the wave function.
  Here we perform
  the auxiliary-field quantum Monte Carlo calculations 
  in the case of small
  systems at finite temperature
  using the complex Langevin method (CLM), which has been successfully
  applied to various interesting systems with the sign problem over the decade.
  In particular, we show the existence of a parameter region in which
  the validity criterion for the CLM is satisfied and 
  the expectation value of the energy obtained by exact diagonalization is correctly reproduced.
  Thus the CLM can be a complementary approach to the variational method
  for large systems.
\end{abstract}
\setcounter{footnote}{0}
\end{titlepage}

\tableofcontents

\section{Introduction}

The nuclear shell model is one of the most powerful theoretical models
that describe not only low-lying energy spectra of nuclei 
including those near the neutron drip line \cite{otsuka_RMP2020,Caurier_RMP,ntsunoda_Nature}
but also thermal nuclear properties \cite{Alhassid_PRC68}.
%%In the framework of the shell-model calculation,
In this model, a frozen inert core is assumed and active particles
in the valence orbits are treated by the so-called model space.
The quantum many-body problem of the active particles
is then solved by diagonalizing the Hamiltonian matrix
in the subspace spanned by a huge number of Slater determinants,
each of which represents how the active particles occupy the single-particle states.
However, the dimension of the Hamiltonian matrix increases
combinatorially as the active particles and the single-particle states increase in number.
While the conventional Lanczos diagonalization method can handle the system 
with the so-called $M$-scheme for dimension up to $O(10^{11})$ \cite{kshell},
the
explosive
%%factorial
growth of the dimension hampers
the shell-model calculations,
%% studies
in particular, in the heavy-mass region.

In order to circumvent the problem of the explosive growth of the dimension, 
various efforts have been made. (See, for instance, Refs.~\cite{OTSUKA2001319,VAMPIR_review}.)
In 1990s, the auxiliary-field Monte Carlo (AFMC) method was introduced in the
shell-model calculations.
Based on this approach, which was named the shell model Monte Carlo (SMMC) method,
$pf$-shell nuclei were studied successfully \cite{KOONIN19971}.
However, practical applications of the method turn out to be severely restricted
by the notorious sign problem.
%%restricts the practical application of the AFMC to fermion many-body problems strictly.
In particular, a realistic effective interaction in the shell model,
which is given, for instance, by the $G$-matrix theory
or by the valence-space in-medium similarity renormalization group (VS-IMSRG)
%%method \cite{Tsukiyama:2012sm,VSIMSRG_review}, cannot be used directly 
method \cite{VSIMSRG_review}, cannot be used directly 
because of the sign problem.
In order to circumvent this problem, one can prepare some artificial interaction
without the sign problem and make an extrapolation \cite{PhysRevLett.72.613},
but such an extrapolation procedure often causes large uncertainties.
This led to the common use of 
%%For this reason,
a schematic interaction without the sign problem instead \cite{Langanke_68Ni_PRC67,OAN13Nu}.
%%is commonly used \cite{Langanke_68Ni_PRC67,OAN13Nu}.
As another attempt,
%%to evade the sign problem, 
the constrained-path approximation was tested
in $sd$- and $pf$-shell nuclei \cite{cpqmc_sm_PRL}.

On the other hand, various attempts have been made to obtain approximate wave functions
using
%%various
variational approaches such as
%%were attempted to obtain the
%%, including
conventional particle-hole truncations,
the VAMPIR method \cite{vampir_ppnp} and 
%%[K. Schmid, Prog. Part. Nucl. Phys. 46, 145 (2001)], and
the Monte Carlo shell model \cite{Shimizu_2017}.
%% [N. Shimizu, T. Abe, M. Honma, T. Otsuka, T. Togashi, Y. Utsuno and T. Yoshida, Phys. Scr. 92, 063001 (2017)].
While these methods are indeed
%%convenient and
useful in practical applications,
they only 
%%these methods
provide us with the upper bound or the extrapolated value
within the variational ansatz,
and furthermore, they cannot be used for finite-temperature physics.
Thus, it is still desirable to develop
%%the development of
quantum Monte Carlo techniques, which do not
have these problems of 
%%rely on
the variational approaches.
%%not relying on the variational approach is still desired.

In this paper, we attempt to solve the sign problem in the shell-model calculations
with realistic interactions
by the complex Langevin method (CLM) \cite{Parisi:1984cs,Klauder:1983sp}.
(See Ref.~\cite{Adami_2001} for an earlier attempt using the Lipkin model.)
%%In particular,
While the method is known to have the wrong convergence problem,
the conditions for the correct convergence have recently been
clarified \cite{Aarts:2009dg,Aarts:2009uq,Aarts:2011ax,Seiler:2012wz,Nishimura:2015pba,Nagata:2015uga,Nagata:2016vkn,Ito:2016efb,Aarts:2017vrv,Scherzer:2018hid}
and various new techniques to satisfy these conditions have been proposed.
Based on these new developments, the CLM has been applied successfully to lattice QCD
at finite density~\cite{Sexty:2013ica,Aarts:2014bwa,Fodor:2015doa,%%
  %%Sinclair:2015kva,Sinclair:2016nbg,Sinclair:2017zhn,Sinclair:2018rbk,
  Nagata:2018mkb,%%
  %%Ito:2018jpo,Tsutsui:2018jva,Tsutsui:2019gwn,
  Kogut:2019qmi,Sexty:2019vqx,%%
  %%Sinclair:2019ysx,
  Tsutsui:2019suq,Scherzer:2020kiu,Ito:2020mys,Tsutsui:2025jez}.
(See Refs.~\cite{Berger:2019odf,Attanasio:2020spv,Nagata:2021ugx} for recent reviews.)
See also Ref.~\cite{YH15Co} for an application to the Hubbard model.
%%and it was also applied to quantum many-body problems such as 
%% Here we apply the CLM to the nuclear shell-model calculations 
%% with realistic interactions.
  %% In particular, we show the existence of a parameter region in which
  %% the validity criterion for the CLM is satisfied and 
  %% the expectation value of the energy obtained by exact diagonalization is correctly reproduced.
  %% % by properly formulating the model so that the overlap problem does not appear.
  %%   Thus the CLM can be a complementary approach to the variational method
  %% for large nuclei.  

We discuss two formalisms, which differ in how we treat the projection onto a fixed
number of active particles.
Our results for finite temperature show that the correct expectation value of energy
is obtained only for sufficiently high temperature, which can be understood
by the validity criterion of the CLM.
In order to obtain results at lower temperature, we propose to introduce a tunable
parameter in the interaction terms and to make an extrapolation using the
data obtained in the parameter region in which the validity criterion of the CLM is satisfied.
Thus we conclude that the CLM can be a useful approach to larger nuclei, which is complementary
to the variational method.
  We hope that
  %%the solution to the sign problem is expected to make a breakthrough also in
  this approach is useful also in 
{\it ab initio} nuclear structure theories based
on the configuration interaction method, such as 
the no-core shell-model method \cite{NCSM_review} and 
the VS-IMSRG
%%valence-space in-medium similarity renormalization group
method \cite{VSIMSRG_review}.
%%method \cite{Tsukiyama:2012sm,VSIMSRG_review}.

This paper is organized as follows.
In section \ref{sec:review},
we briefly review how the CLM works and discuss the validity criterion of the method.
In section \ref{sec:app-shell-model},
we discuss how we apply the CLM to the nuclear shell model,
and present our results in section \ref{sec:results}.
Section \ref{sec:summary} is devoted to a summary and discussions.
In Appendix \ref{sec:appendix}, we discuss how to
implement the projection onto a fixed number of active particles
in one of the formalisms discussed in this paper.
%%for the implementation of the canonical formalism in the CLM.
In Appendix \ref{sec:RLM}, we present some results for the case without the sign problem
to confirm that the real Langevin simulation works in that case.
In Appendix \ref{sec:g-extrapolation}, we discuss the $g$-extrapolation
as an alternative way to perform extrapolations from the region
in which correct results are available.
%% which is called 
%%
%%present some results of the so-called $g$-extrapolation for comparison.
%%
%% , where the data in the $g<0$ region
%% are obtained correctly due to the absence of the sign problem
%% and $g=1$ corresponds to the target Hamiltonian.

\section{Brief review of the complex Langevin method}
\label{sec:review}

In this section, we briefly review 
how the complex Langevin method works.
For that, let us consider the integral
\begin{align}
 \langle \mathcal{O}(X) \rangle
 =\frac{1}{Z}\int dX\, \mathcal{O}(X) e^{-S(X)} \  ,
 \qquad
 Z=\int dX\, e^{-S(X)} \  ,
\end{align}
where $X$ represents a set of dynamical variables and $S(X)$ represents its function.
If $S(X)$ is a real function, $e^{-S(X)}/Z$ can be interpreted as a probabilistic weight
so that one can numerically evaluate the integral by stochastic methods.

One of such stochastic methods is the so-called (real) Langevin method,
which is based on the equivalence between the Fokker-Planck equation and the Langevin equation.
Let us consider the following Fokker-Planck equation
\begin{align}
 \pder{}{t}P(X;t)
 =\pder{}{X}\left(  \pder{}{X}
 % +\pder{S}{X}
 -v(X)  \right) P(X;t)  \ ,
 \label{FPeq}
\end{align}
where $P(X;t)$ is the probability distribution for $X$ at time $t$
and $v(X)$ is the drift term defined by $v(X)=-\pder{S}{X}$.
If $S$ is real and non-singular,
the solution to the Fokker-Planck equation 
uniquely asymptotes to
%%the weight
$e^{-S(X)}/Z$ in the large $t$ limit.
% after sufficiently long time has elapsed.
It is well known that this Fokker-Planck equation is
%%Then,
equivalent to the Langevin equation 
\begin{align}
 \der{x(t)}{t}
 =v(x(t)) % -\pder{S}{x}(x(t))
 +\eta(t) \  ,
 \label{RLangevin-eq}
\end{align}
where $\eta(t)$ is the white noise that satisfies
\begin{align}
 \langle \eta(t) \rangle_\eta = 0 \ ,
 \qquad
 \langle \eta(t)\eta(t') \rangle_\eta = 2\delta(t-t') \  ,
 \label{white-noise-cond}
\end{align}
with $\langle \cdots \rangle_\eta$ being an average with respect to the white noise,
which can be generated by the Gaussian distribution
$\propto\exp[-\frac{1}{4}\int dt\,\eta(t)^2]$.
The solution $x(t)$ to the Langevin equation 
then corresponds to the solution $P(X;t)$
%%provides the probability distribution in
to the Fokker-Planck equation through
$P(X;t)=\langle\delta(X-x(t))\rangle_\eta$.
Thus
%%the Langevin equation generates the desired probability distribution $e^{-S(X)}/Z$
%%after sufficiently long $t$;
%%namely
one just needs to solve the Langevin equation for a long enough $t$
%%to realize the distribution.
to realize the desired probability distribution $e^{-S(X)}/Z$.

The complex Langevin method is an extension of the real Langevin method
to a system with a complex-valued $S(x)$.
%a complex weight. Let us now consider a complex-valued $S(x)$.
In that case, it is natural to consider a complex version of the Langevin equation
\begin{align}
 \der{z(t)}{t}=v(z(t))%-\pder{S}{z}(z(t))
 +\eta(t) \  ,
 \label{CLangevin-eq}
\end{align}
where $z(t)$ represents a set of complexified dynamical variables,
$v(z)=-\pder{S}{z}$ represents the drift term,
and $\eta(t)$ represents the real-valued white noise that satisfies \eqref{white-noise-cond}.
The equivalent Fokker-Planck equation can be obtained as
\begin{align}
 \pder{}{t}\mathcal{P}(Z,\bar Z;t)
 &=\left\{ \pder{}{Z}\left(
 \pder{}{Z}
 -v(Z) %+\pder{S}{Z}
 \right) 
 +\pder{}{\bar Z}\left(
 \pder{}{\bar Z}
 -\overline{v(Z)} %+\overline{\pder{S}{Z}}
 \right) 
 +2\pder{^2}{Z\partial \bar Z}
 \right\} \mathcal{P}(Z,\bar Z;t)
 \nonumber \\
 &=\left(
 \pder{^2}{X^2}
 % +\pder{}{X}\operatorname{Re}\left[\pder{S}{Z}\right]
 -\pder{}{X}\operatorname{Re}\left[v(Z)\right]
 % +\pder{}{Y}\operatorname{Im}\left[\pder{S}{Z}\right]
 -\pder{}{Y}\operatorname{Im}\left[v(Z)\right]
 \right) \mathcal{P}(Z,\bar Z;t) \ ,
 \label{CFP-eq}
\end{align}
where $X$ and $Y$ are the real and imaginary parts of $Z$, respectively,
and $\mathcal{P}(Z,\bar Z;t)$ is the probability distribution for $Z$ and $\bar Z$
at time $t$.
% via $Z=X+iY$.
If the following equation holds
\begin{align}
 \int dX\, P(X;t)\mathcal{O}(X)
 =\int dZ\, d\bar Z\, \mathcal{P}(Z,\bar Z;t)\mathcal{O}(Z) \  ,
 \label{ev-relation}
\end{align}
where $P(X;t)$ is a complex-valued function that satisfies \eqref{FPeq}
with the complex-valued $S(X)$,
a solution to the complex Langevin equation \eqref{CLangevin-eq}
gives the expectation value of observables with the complex weight $P(X;t)$.

Let us note first that, unlike the real Langevin case, 
$P(X;t)$ does not necessarily converge to $e^{-S}/Z$.
This is not a problem, however, as far as
$\mathcal{P}(Z,\bar Z;t)$
%%$P(X;t)$
that satisfies \eqref{ev-relation}
converges uniquely to some finite function in the $t\rightarrow \infty$ limit
%%approaches infinity 
since in that case, one can prove that 
%%because it was found in Ref.~\cite{Nishimura:2015pba} that
the $t\to\infty$ limit of $P(X;t)$ should be $e^{-S}/Z$ \cite{Nishimura:2015pba}.

Thus the question boils down to whether \eqref{ev-relation} holds,
and if so, under which condition.
By formal integration of the Fokker-Planck equation \eqref{CFP-eq},
the right-hand side of \eqref{ev-relation}
can be rewritten as \cite{Aarts:2009uq,Aarts:2011ax}
\begin{align}
 \int dZ\, d\bar Z\, \mathcal{P}(Z,\bar Z;t)\mathcal{O}(Z)
 =\int dZ\, d\bar Z\, \{ e^{tL^{T}}\mathcal{P}(Z,\bar Z;0) \}\mathcal{O}(Z) \  ,
\end{align}
where $L$ is the Langevin operator for \eqref{CLangevin-eq} given by
\begin{align}
 L=\pder{^2}{X^2}
 % -\operatorname{Re}\left[\pder{S}{Z}\right] \pder{}{X}
 +\operatorname{Re}\left[ v(Z) \right] \pder{}{X}
 % -\operatorname{Im}\left[\pder{S}{Z}\right] \pder{}{Y}
 +\operatorname{Im}\left[ v(Z) \right] \pder{}{Y} \  ,
\end{align}
and its transpose $L^\top$ reads
%the Fokker-Planck operator for \eqref{CFP-eq},
\begin{align}
 L^\top=\pder{^2}{X^2}
 % +\pder{}{X}\operatorname{Re}\left[\pder{S}{Z}\right]
 -\pder{}{X}\operatorname{Re}\left[ v(Z) \right]
 % +\pder{}{Y}\operatorname{Im}\left[\pder{S}{Z}\right]
 -\pder{}{Y}\operatorname{Im}\left[ v(Z) \right] \  .
\end{align}
%% Then the equality \eqref{ev-relation} can be derived if one assumes that
%%needs to
Let us here assume that
there is no boundary term
in the following integration by parts
\begin{align}
 \int dZ\, d\bar Z\, \{ e^{tL^{T}}\mathcal{P}(Z,\bar Z;0) \}\mathcal{O}(Z)
 =\int dZ\, d\bar Z\, \mathcal{P}(Z,\bar Z;0)\{ e^{tL}\mathcal{O}(Z) \} \  ,
 \label{int-by-parts}
\end{align}
where the right-hand side is well-defined in the $t\rightarrow \infty$ limit.
We also assume that the operator $\mathcal{O}(Z)$ is a holomorphic function of $Z$
and that the initial condition is $\mathcal{P}(Z,\bar Z;0)=P(X;0)\delta(Y)$.
Under these assumptions, one can derive \eqref{ev-relation} as
\begin{align}
 % \int dZ\, d\bar Z\, \{ e^{tL^{T}}\mathcal{P}(Z,\bar Z;0) \}\mathcal{O}(Z)
 \int dZ\, d\bar Z\, \mathcal{P}(Z,\bar Z;0)\{ e^{tL}\mathcal{O}(Z) \}
 &=\int dX\, P(X;0)\{ e^{tL_0}\mathcal{O}(X) \}
 \nonumber \\
 &=\int dX\, \{ e^{tL_0^\top}P(X;0) \}\mathcal{O}(X)
 \nonumber \\
 &=\int dX\, P(X;t) \mathcal{O}(X) \  ,
\end{align}
where
we have used
$L\mathcal{O}(Z)|_{Y=0}=L_0\mathcal{O}(X)$ for holomorphic $\mathcal{O}(Z)$
and defined
\begin{align}
 L_0=\pder{^2}{X^2}
 +v(X) %-\pder{S}{X}
 \pder{}{X} \ ,
 \qquad
 L_0^\top=\pder{^2}{X^2}
 -\pder{}{X}
 v(X) %(-\pder{S}{X})
  \ .
\end{align}

Thus the key point in justifying the CLM is the validity of \eqref{int-by-parts},
which depends on the model as well as its parameter region.
There are some criteria \cite{Aarts:2011ax,Nagata:2016vkn,Salcedo:2016kyy,Scherzer:2018hid}
for the absence of the boundary term in \eqref{int-by-parts}.
%%proposed in Refs.~\cite{Aarts:2011ax,Nagata:2016vkn,Salcedo:2016kyy,Scherzer:2018hid}.
In this paper, we use the one proposed in
Ref.~\cite{Nagata:2016vkn},
which is a practical and convenient criterion based on
a sufficient condition for \eqref{int-by-parts}.
This criterion focuses on the probability distribution of the magnitude of the drift term $v(z)$
generated by $\mathcal{P}(Z,\bar Z;t)$.
If the probability distribution falls off exponentially or faster at large magnitude,
the formal expression of the right-hand side in \eqref{int-by-parts}
is well defined and the boundary term in the integration by parts vanishes.

In practice,
we simulate the model with the discretized complex Langevin equation
\begin{align}
 z(t+\iDelta t)
 =z(t)+ v(z(t))\iDelta t
 +\tilde\eta(t)\sqrt{\iDelta t} \  ,
\end{align}
where $\tilde\eta(t)$ is generated
by the Gaussian distribution $\propto\exp[-\frac{1}{4}\tilde\eta(t)^2]$,
monitoring the maximum value of the magnitude of the drift term.
The expectation value of an observable is obtained by the average of
the observable measured at each Langevin time after thermalization.
We check the validity criterion by making a histogram for the maximum value of the
magnitude of the drift term.

Even when the CLM is supposed to work,
it may fail because of 
some instability in the time-evolution by the discretized complex Langevin equation.
This can be controlled by using an adaptive step-size for the discretized time-evolution,
which depends on the magnitude of the drift term \cite{Aarts:2009dg}.
In this paper, we choose the step-size $\iDelta t$ such that
$\iDelta t=\min(\iDelta t_0, \iDelta t_0\frac{v_0}{\max(|v(z(t))|)})$,
where $\iDelta t_0$ and $v_0$ are positive parameters that we set at the beginning of a simulation.
Since the model we simulate has more than one dynamical variables,
we use $\max(|v(z(t))|)$ for the adaptive step-size, 
which is the maximum value of the absolute value of the elements of the drift term.
Note that one needs to take the adaptive step-size into account
when one makes
%%in making
the drift histogram for the criterion.

\section{Applying the CLM to the shell model}
\label{sec:app-shell-model}

In this section we apply the CLM reviewed in the previous section to
the nuclear shell-model calculations.
After defining the shell model, we discuss how it can be investigated by the CLM.
%%and present our results. 

%%\subsection{Definition of the shell model}
%%We apply the CLM to nuclear shell-model calculations.
Let us consider a many-body problem of the active particles
in the model space, in which
the number of the single-particle states
%%of the model space 
and that of the active particles
%%inside the model space
are $N_s$ and $N_v$, respectively.
The many-body shell-model Hamiltonian up to the two-body interaction
is written in the density-decomposed form as \cite{KOONIN19971}
%
% In nuclear shell-model calculations,
% we solve the many-body problem of active particles
% in the model space. 
% The numbers of the single-particle states of the model space 
% and the active particles inside the model space
% are referred to as $N_s$ and $N_v$, respectively.
% In general, the two-body shell-model Hamiltonian is 
% written in the density-decomposed form 
% \cite{KOONIN19971} as 
%
% In the nuclear shell model, 
% once a model space, or equivalently a shell, is specified, 
% the number of single-particle states, denoted by $N_s$, is determined.
% Then, many-body states in a shell-model system 
% are constructed by creation operators of the single-particle states.
% Some shell-model Hamiltonians that reproduce experimental data 
% are known and perturbatively constructed.
% If one truncates such a many-body Hamiltonian up to two-body interaction terms,
% it can be written in the form of
\begin{align}
 \hat H
 =\sum_{i,j=1}^{N_s}T_{ij} \, \hat c_i^\dagger \, \hat c_j
 +\frac{1}{2}\sum_{\alpha} V_\alpha \left( \hat O_\alpha \right)^2
\  ,
 \label{Hamiltonian}
\end{align}
where 
$\hat c_i^\dagger$ represents the creation operator of a single-particle state
% $T_{ij}$ is a single-particle energy,
and $\hat O_\alpha$ represents the density operator of the form $\hat c^\dagger \hat c$.
We have introduced $T_{ij}$, which denotes the strength of the one-body term, 
and $V_\alpha$, which denotes the strength of the two-body interaction.

We simulate this theory at finite temperature by using the standard imaginary-time formalism,
where the imaginary time $\beta$, which corresponds to the inverse temperature,
is sliced up into $n_\beta$ segments.
%%In order to perform numerical simulations of this theory, 
%%we discretize 
%%We denote the number of the time slices by $n_\beta$,
The partition function for a fixed number $N_v$  of valence nucleons is written as
\begin{align}
 Z= \Tr_{N_v} [(e^{-\iDelta\beta \hat H})^{n_\beta}] \  ,
\end{align}
where $\iDelta\beta=\beta/n_\beta$ and
$\Tr_{N_v}$ is a trace in the Fock space with $N_v$ particles.
Here we use the Hubbard-Stratonovich transformation at each imaginary time
to linearize the Hamiltonian with respect to $\hat O_\alpha$
%%in terms of only single-particle operators
by introducing new auxiliary scalar fields $\sigma_{\alpha n}$,
where $n$ runs from 1 to $n_\beta$.
Then the partition function becomes \cite{KOONIN19971}
\begin{align}
  Z&\simeq\int [d\sigma] \, 
  (\Tr_{N_v} \hat U)\, e^{-\frac{1}{2}\iDelta\beta \sum_{\alpha n} |V_{\alpha}| \sigma_{\alpha n}^2} \  ,
\label{def-Z-canonical}
  \\
   \hat U&=e^{-\iDelta\beta \hat h_{n_\beta}}e^{-\iDelta\beta \hat h_{n_\beta-1}}\cdots e^{-\iDelta\beta \hat h_{1}} \ , \\
%% \nonumber \\
 \hat h_{n}&=\sum_{i,j}T_{ij}\hat c_i^\dagger \hat c_j
 +\sum_{\alpha} s_{\alpha} V_\alpha \sigma_{\alpha n} \hat O_\alpha \  ,
 \label{def-h-hat}
\end{align}
%where $s_{\alpha}$ is a phase factor that takes 1 for $V_\alpha<0$ and $i$ for $V_\alpha>0$.
where $s_{\alpha}=1$ for $V_\alpha<0$ and $s_{\alpha}=i$ for $V_\alpha>0$.
The integration measure for $\sigma$ in \eqref{def-Z-canonical} is defined by
$[d \sigma] = \prod_{\alpha, n} \left( d\sigma_{\alpha n}
\sqrt{\frac{\Delta \beta |V_{\alpha}|}{2\pi}} \right)$.
The expectation value of an observable $\mathcal{O}$
%% the path integral to compute an observable, $\mathcal{O}$, 
with fixed $N_v$ can be written as
\begin{align}
 \langle \mathcal{O} \rangle_{N_v}
 =\frac{1}{Z}\int [d\sigma] \frac{\Tr_{N_v} \hat U\mathcal{\hat O}}{\Tr_{N_v} \hat U}
 (\Tr_{N_v} \hat U)\, e^{-\frac{1}{2}\iDelta\beta \sum_{\alpha n} |V_{\alpha}| \sigma_{\alpha n}^2} \  .
 \label{fixed-Nv-path-integral}
\end{align}
%% where
%% \begin{align}
%% \end{align}
Let us call this formalism \eqref{fixed-Nv-path-integral} the canonical formalism in this paper. 
% $N_v$-fixed formalism

In fact, the canonical formalism is not very convenient
since $\Tr_{N_v}\hat U$ takes different forms
for different $N_v$, and it becomes more and more complicated as $N_v$ increases.
%%
%% provide different forms of 
%% the explicit form of the drift term depends on the value of $N_v$.
%%
%% if we simulate systems with different values of 
%% the number of valence nucleons, %in equation \eqref{fixed-Nv-path-integral} 
%% $N_v$.
%% Since different $N_v$ provide different forms of $\Tr_{N_v}\hat U$,
%% the explicit form of the drift term depends on the value of $N_v$.
%% Therefore, one needs to prepare a different code for each value of $N_v$.
%%
% In this section,
% we propose and apply a different approach 
%%To resolve this issue,
From this point of view, it is beneficial to rewrite
the path integral into the grand canonical ensemble
with the number projection \cite{KOONIN19971}.
Using the projection onto a fixed $N_v$ given by
\begin{align}
 \delta(\hat N-N_v)=
 \int_0^{2\pi} \frac{d\phi}{2\pi}\,
 e^{(\beta \mu +i\phi)(\hat N-N_v)} \  ,
\end{align}
we rewrite \eqref{fixed-Nv-path-integral} as
\begin{align}
 \langle \mathcal{O} \rangle_{N_v}
 &=\frac{1}{Z}\int [d\sigma] \int \frac{d\phi}{2\pi}
 \Tr\left[ e^{(\beta \mu +i\phi)(\hat N-N_v)}\hat U\mathcal{\hat O} \right]
 e^{-\frac{1}{2}\iDelta\beta \sum_{\alpha n} |V_{\alpha}| \sigma_{\alpha n}^2} \  , \nonumber \\
 Z &= \int [d\sigma] \int \frac{d\phi}{2\pi}
 \Tr\left[ e^{(\beta \mu +i\phi)(\hat N-N_v)}\hat U \right]
 e^{-\frac{1}{2}\iDelta\beta \sum_{\alpha n} |V_{\alpha}| \sigma_{\alpha n}^2}  \ ,
\end{align}
where $\hat N$ is the number operator of valence nucleons
and $\mu$ is the chemical potential.
Here $\Tr$ is a trace in the entire Fock space of the model space,
which sums over all possible $N_v$.
In the CLM, $\mu$ is redundant since $\phi$ is a complex dynamical variable,
and therefore we set $\mu=0$ from now on.
It is straightforward to rewrite 
the weight $\Tr e^{i\phi(\hat N-N_v)}\hat U$
for the grand canonical ensemble by single-particle operators as
\begin{align}
 \Tr e^{i\phi(\hat N-N_v)}\hat U
 =e^{-i\phi N_v}\det\left( 1+e^{i\phi} U \right) \  ,
\end{align}
where $U$ is a one-body matrix defined by $U_{ij}=(\exp[-\beta h^{\rm (tot)}])_{ij}$
with the matrix $h^{\rm (tot)}$ defined by
$\hat U = \exp[-\beta\sum_{i,j}h^{\rm (tot)}_{ij}\hat c_i^\dagger \hat c_j]$.
Thus, the expectation value in the case of a one-body operator $\mathcal{\hat O}$
can be written as
%%becomes
\begin{align}
 \langle \mathcal{O} \rangle_{N_v}
 &=\frac{1}{Z}\int [d\sigma] \int \frac{d\phi}{2\pi} \, e^{-i\phi N_v}
 \det\left( 1+e^{i\phi} U \right)
 e^{-\frac{1}{2}\iDelta\beta \sum_{\alpha n} |V_{\alpha}| \sigma_{\alpha n}^2}
 % \nonumber \\
 % &\qquad\qquad
 \; \tr\left[ \mathcal{U}\mathcal{O} \right] \  ,
 \nonumber \\
 Z&=\int [d\sigma] \int \frac{d\phi}{2\pi} \, e^{-i\phi N_v}
 \det\left( 1+e^{i\phi} U \right)
 e^{-\frac{1}{2}\iDelta\beta \sum_{\alpha n} |V_{\alpha}| \sigma_{\alpha n}^2} \  ,
 \label{grand-canonical-path-integral}
\end{align}
where $\tr$ is the matrix trace for $N_s\times N_s$ matrices and we have defined
\begin{align}
 \mathcal{U}
 =\left( 1+ e^{i\phi} U \right)^{-1}
 e^{i\phi} U \  .
\end{align}
Let us call \eqref{grand-canonical-path-integral} the grand-canonical formalism in this paper.

This formula \eqref{grand-canonical-path-integral} can be generalized to the
case in which $\mathcal{\hat O}$ is a product of
one-body operators $\mathcal{\hat O}_1,\cdots,\mathcal{\hat O}_n$ using
%%as a polynomial of one-body operators.
%% The factor $\tr[\mathcal{UO}]$ in Eq.~\eqref{grand-canonical-path-integral} differs
%% depending on the degree of $\mathcal{O}$ as a polynomial of one-body operators.
%% In general, it can be computed 
%% for the product of one-body operators $\mathcal{\hat O}_1,\cdots,\mathcal{\hat O}_n$ by
%% 
\begin{align}
 \Tr [e^{i\phi(\hat N-N_v)}\hat U \mathcal{\hat O}_1\cdots \mathcal{\hat O}_n]
 % =\pder{}{\epsilon_1}\cdots\pder{}{\epsilon_n}
 % \Tr [e^{i\phi(\hat N-N_v)}\hat U_{\epsilon_1,\cdots,\epsilon_n}]_{\epsilon_1,\cdots,\epsilon_n\to 0}
 =e^{-i\phi N_v}
 \left.
 \pder{}{\epsilon_1}\cdots\pder{}{\epsilon_n}
\det\left( 1+e^{i\phi} U_{\epsilon_1,\cdots,\epsilon_n} \right)
\right|_{\epsilon_1 = \dots = \epsilon_n = 0}
\  ,
\end{align}
where $\hat U_{\epsilon_1,\cdots,\epsilon_n}=\exp[-\beta\sum_{i,j}h^{\rm (tot)}_{ij}\hat c_i^\dagger \hat c_j]\exp[\epsilon_1\mathcal{\hat O}_1]\cdots\exp[\epsilon_n\mathcal{\hat O}_n]$
corresponding to a one-body matrix $(U_{\epsilon_1,\cdots,\epsilon_n})_{ij}
=(\exp[-\beta h^{\rm (tot)}]\exp[\epsilon_1\mathcal{O}_1]\cdots\exp[\epsilon_n\mathcal{O}_n])_{ij}$.
For instance, the energy \eqref{Hamiltonian}, which involves two-body operators,
can be written as
\begin{align}
 \langle H \rangle_{N_v}
 &=\frac{1}{Z}\int [d\sigma] \int \frac{d\phi}{2\pi} \, e^{-i\phi N_v}
 \det\left( 1+e^{i\phi} U \right)
 e^{-\frac{1}{2}\iDelta\beta \sum_{\alpha n} |V_{\alpha}| \sigma_{\alpha n}^2}
 \nonumber \\
 &\qquad\quad
 \times\left(
 \tr\left[\mathcal{U}T
 \right]
 +\frac{1}{2}\sum_{\alpha}V_\alpha\left\{
 \left( \tr\left[\mathcal{U}O_\alpha
 \right] \right)^2
 +\tr\left[\mathcal{U}O^2_\alpha
 \right]
 -\tr\left[\mathcal{U}O_\alpha
 \mathcal{U}O_\alpha
 \right]
 \right\}
 \right) \ .
\end{align}
The standard procedure in the literature is
to reduce the integration over $\phi$ to a summation over $N_s$ integers \cite{KOONIN19971}.
In the CLM, however, we leave the integration over $\phi$ as it is
and treat $\phi$ as another integration variable.
%% undone because it spoils the form of the observable 
%% and makes it in the form of a numerator of a trace over a denominator of another trace,
%% which would lead us to the overlapping problem.

% \subsection{Drift terms for the grand-canonical formalism}

In our simulations,
we change the normalization of the variables as
$\sigma_{\alpha n}=\frac{\tilde{\sigma}_{\alpha n}}{\sqrt{\iDelta\beta |V_{\alpha}|}}$.
%%we write down explicit expressions of the drift terms in the grand-canonical formalism.
%%First, just for simplicity, let us
The effective action then becomes
\begin{align}
 S=\sum_{\alpha,n}\frac{\tilde{\sigma}_{\alpha n}^{2}}{2}
 -\ln\det\left[
 1+e^{i\phi}U
 \right]
 +iN_v\phi \ .
\end{align}
Using this effective action, the Langevin equations are obtained as
%%governed by this effective action are written as 
\begin{align}
 \der{\tilde{\sigma}_{\alpha n}}{t}
 =-\pder{S}{\tilde{\sigma}_{\alpha n}}+\eta_{\alpha n}(t) ,
 \qquad
 \der{\phi}{t}
 =-\pder{S}{\phi}+\eta_\phi(t) \ ,
\end{align}
where $\eta_{\alpha n}$ and $\eta_\phi$ represent the Gaussian noise, which satisfies
$\langle \eta_{\alpha n}(t)\eta_{\beta m}(0) \rangle =2\delta_{\alpha\beta}\delta_{nm}\delta(t)$ and
$\langle \eta_\phi(t)\eta_\phi(0) \rangle =2\delta(t)$.
The drift terms are calculated as
\begin{align}
 \pder{S}{\tilde{\sigma}_{\alpha n}}
 &=\tilde{\sigma}_{\alpha n}-\tr \left[
 \left( 1+e^{i\phi}U \right)^{-1}
 e^{i\phi}\pder{U}{\tilde{\sigma}_{\alpha n}}
 \right]
%% \nonumber \\
 %% &=
 =
 \tilde{\sigma}_{\alpha n}-\tr \left[
 \mathcal{U} U^{-1}\pder{U}{\tilde{\sigma}_{\alpha n}}
 \right]  \ ,
 \nonumber \\
 \pder{S}{\phi}
 &=-i\tr \left[
 \left( 1+e^{i\phi}U \right)^{-1}
 e^{i\phi}U
 \right]
 +iN_v
%% \nonumber \\
 %% &=
 =
 -i\left( \tr\left[ \mathcal{U} \right] -N_v \right)  \ .
\label{eq:drift-terms}
\end{align}
%%We use these expressions to simulate the nuclear shell model.
See Appendix \ref{sec:appendix} for the implementation of the canonical formalism in the CLM.

%%%%%%%%%%%%%%%% Asano estimate %%%%%%%%%
The computational cost for the grand-canonical
formalism\footnote{The computational cost for the canonical formalism
is $O((N_s)^5 n_\beta)$ as well
%%scales in the same manner 
since the calculation of $\pder{U}{\tilde{\sigma}_{\alpha n}}$ appears
in the drift term \eqref{eq:drift-canonical}.}
grows as $O((N_s)^5 n_\beta)$.
The dominant part comes from the calculation of the drift terms \eqref{eq:drift-terms}
in the complex Langevin equation, which contains
the calculations of $U$, $\mathcal{U}$ and $\pder{U}{\tilde{\sigma}_{\alpha n}}$.
The calculation of $U$ involves
matrix multiplications in the imaginary-time direction, which cost
$O((N_s)^3 n_\beta)$.
%%, whereas
The calculation of $\mathcal{U}$ involves taking the inverse
of $N_s \times N_s$ matrices, which costs $O((N_s)^3)$.
The most dominant part comes from the calculation of $\pder{U}{\tilde{\sigma}_{\alpha n}}$
since it involves matrix multiplications at each $\alpha= 1 , \cdots , N_\alpha$,
which costs $O(N_\alpha (N_s)^3 n_\beta )$.
%%, where $N_\alpha$ is the number of values that the index $\alpha$ can take,
Using $N_\alpha=(N_s)^2$, which is the case in our implementation,
the computational cost becomes $O((N_s)^5 n_\beta)$.
%%%%%%%%%%%%%%%

\section{The results of the CLM}
\label{sec:results}
% We show our results obtained by the CLM in this subsection.

%%The main observable we measured is the energy.

In this section, we present our numerical results for the shell model
obtained by the CLM.
We focus on the case with $N_s=20$ in the $pf$ shell, 
considering only valence neutrons for simplicity.
The interaction we use in the Hamiltonian \eqref{Hamiltonian}
is taken from GXPF1A in Ref.~\cite{Honma_2005}.
%% Although we simulated various shell-model systems from $N_s=4$ to $32$,
%% we focus, in this paper, on the case
Throughout this paper, we set the Langevin step-size to
$\iDelta t=5\times 10^{-3}$, which
is found to be small enough for the accuracy required in this paper.

%% Even though one needs to measure the observables 
%% with various values of the Langevin step
%% and to extrapolate them to $\iDelta t\to 0$,
%% we set it to $\iDelta t=5\times 10^{-3}$ throughout this paper
%% because we find this value is small enough for our purpose.

\subsection{Dependence on the temperature}
%%$\beta$

\begin{figure}[H]
 \centering
 \includegraphics[scale=0.8]{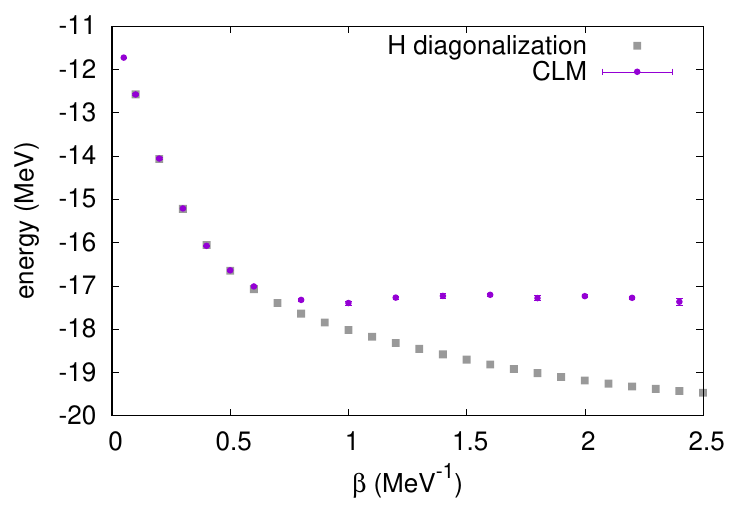}
 \includegraphics[scale=0.6]{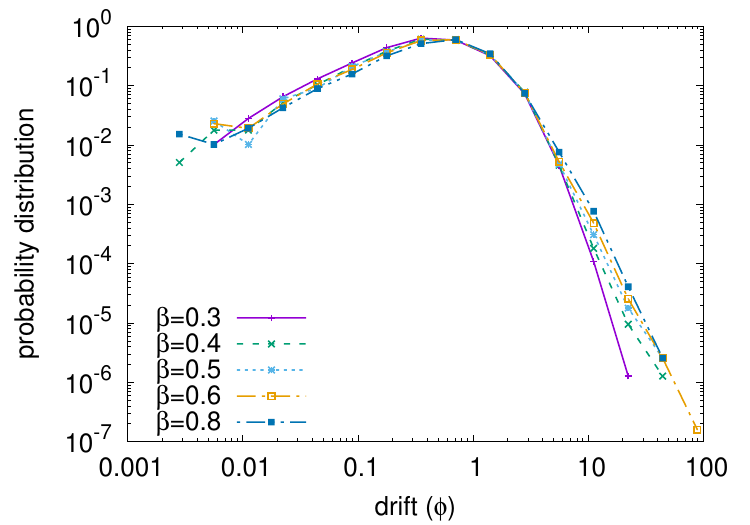}
 \includegraphics[scale=0.6]{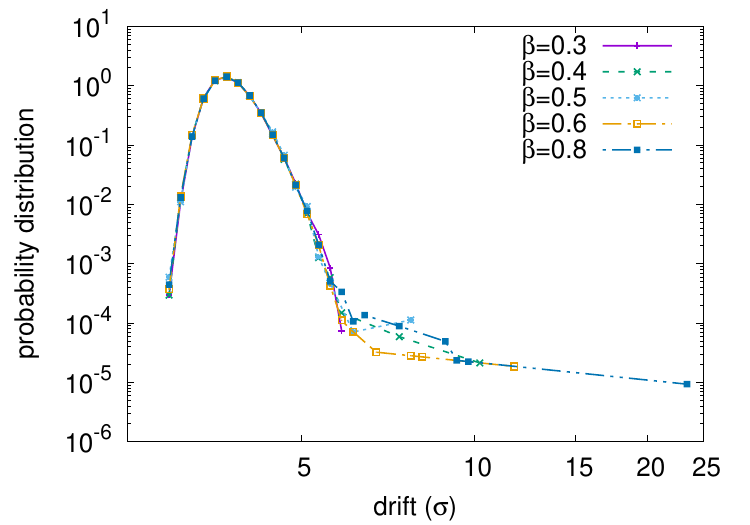}
 \caption{(Top) The expectation value of the
   energy is plotted against the inverse temperature $\beta$
   in the grand-canonical formalism with $N_s=20$, $N_v=2$, $n_\beta=10$.
%%   in the grand-canonical formalism with $N_s=20$, $N_v=2$, $n_\beta=10$.
   The purple points with error bars represent the CLM results,
   whereas the gray points represent the results obtained by the Hamiltonian diagonalization.
(Bottom) The drift histogram for the $\phi$ field (Left) and the $\sigma$ field (Right).}
 \label{fig:NPphi_Ns20_Nt10_Nv2_Energy}
\end{figure}

Let us discuss how our results depend on the temperature.
In Fig.~\ref{fig:NPphi_Ns20_Nt10_Nv2_Energy},
we show our results
%%for the expectation value of the energy
%%Let us first consider the case with
for $N_v=2$ in the grand-canonical formalism with various inverse temperature $\beta$.
The number of steps in the imaginary time direction is set to $n_\beta=10$.
%%In Fig.~\ref{fig:NPphi_Ns20_Nt10_Nv2_Energy} (Top),
In the Top panel, we plot the expectation value of the energy in units of ${\rm MeV}$
against $\beta$ in units of ${\rm MeV}^{-1}$.
We observe good agreement
with the results obtained by the Hamiltonian diagonalization
for $\beta\lesssim 0.6$.
On the other hand, the energy ceases to decrease monotonically for $\beta\gtrsim 0.6$,
which represents an unphysical behavior.
%%In Fig.~\ref{fig:NPphi_Ns20_Nt10_Nv2_Drifts},
%
%In Fig.~\ref{fig:NPphi_Ns20_Nt10_Nv2_Energy} (Bottom)

In the Bottom panel,
we show the histogram of the drift terms.
Since the histogram
%%for the $\phi$ field
for $\beta=0.3$ seems to fall off exponentially
for both $\phi$ and $\sigma$ fields,
we can safely say that the CLM for $\beta\lesssim 0.3$ is justified,
which is consistent with the agreement
with the results of the Hamiltonian diagonalization in that region.
On the other hand,
the drift histogram for the $\sigma$ field
does not show a fast fall-off for $\beta \gtrsim 0.4$,
which suggests that the agreement
with the results of the Hamiltonian diagonalization 
in the $0.4 \lesssim \beta \lesssim 0.6$ is rather accidental.
%%since the criterion is a sufficient condition.

\begin{figure}[H]
 \centering
 \includegraphics[scale=0.8]{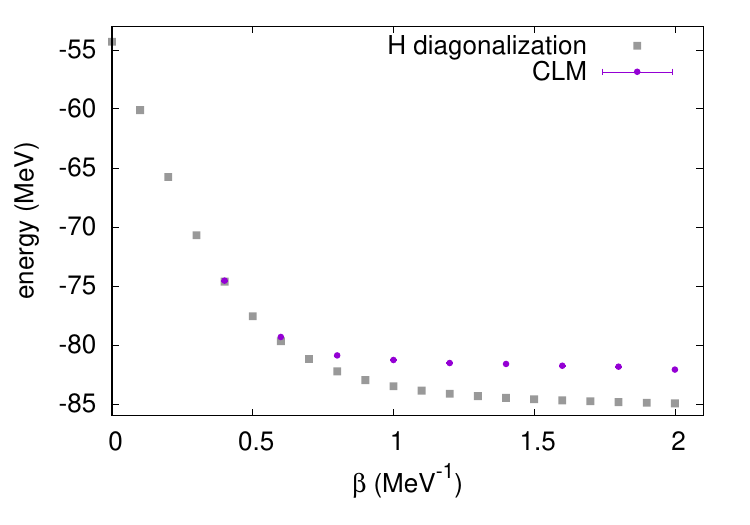}
 \includegraphics[scale=0.6]{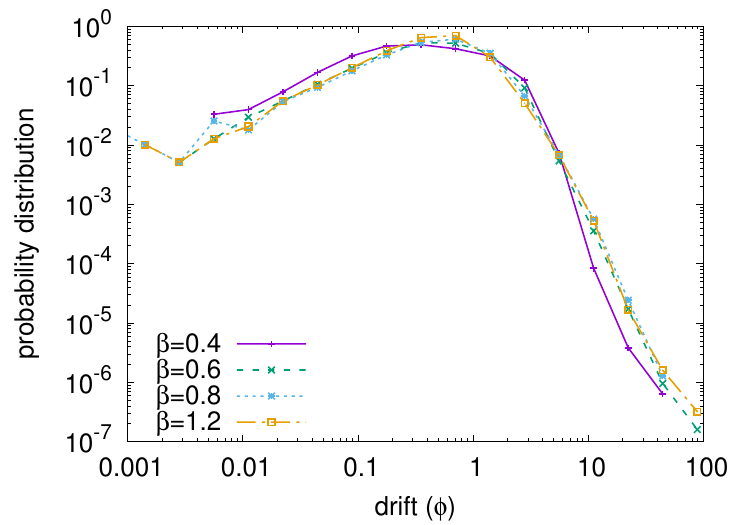}
 \includegraphics[scale=0.6]{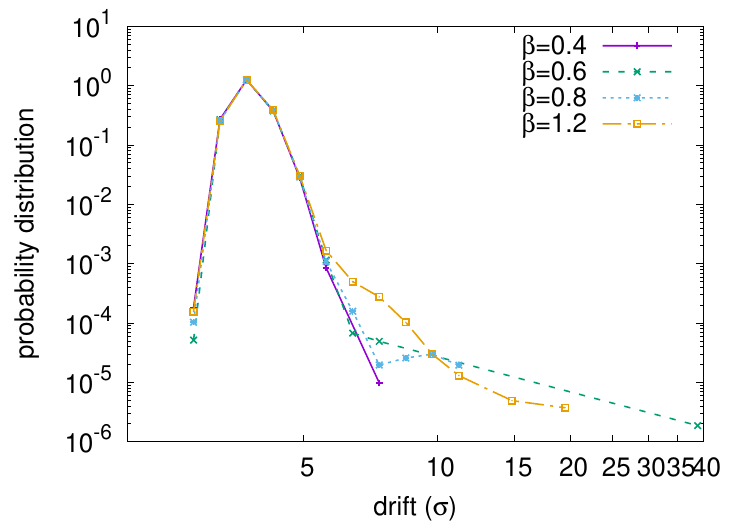}
 \caption{Similarly to Fig.~\ref{fig:NPphi_Ns20_Nt10_Nv2_Energy},
   we present the results for the grand-canonical formalism
   with $N_s=20$, $N_v=10$, $n_\beta=10$.}
 \label{fig:NPphi_Ns20_Nt10_Nv10_Energy}
\end{figure}

Let us recall that
the grand-canonical formalism allows us to obtain results for different $N_v$
in a straightforward manner unlike the canonical formalism.
In Fig.~\ref{fig:NPphi_Ns20_Nt10_Nv10_Energy}
we show our results for $N_v=10$ with $n_\beta=10$.
%% Our CLM results for $N_v=10$ with $n_\beta=10$ are shown in 
%% Fig.~\ref{fig:NPphi_Ns20_Nt10_Nv10_Energy}
%% and Fig.~\ref{fig:NPphi_Ns20_Nt10_Nv10_Drifts}.
We observe good agreement with the result of the Hamiltonian diagonalization
for $\beta \lesssim 0.4$, which is consistent with the result of the drift histogram.

As we have seen above,
the CLM
works only at sufficiently small $\beta$ (high temperature),
%%does not work at low temperature (large $\beta$),
at least for the $N_s=20$ case with valence neutrons.
%% due to the singular drift problem.
However, if we are interested in the ground state energy,
we have to take the $\beta \rightarrow \infty$  limit.
Since the result for finite $\beta$ is a weighted sum over
the low-lying energy levels $E_n$ with the Boltzmann weight $e^{-\beta E_n}$, 
it is not straightforward to extract the energy levels $E_n$ from the observed
$\beta$ dependence of the energy expectation values.
%
%% Since the reason why we are interested in the $\beta$-dependence of energy
%% is to obtain information of energy spectrum though it is almost impossible to read off the detailed energy levels,
%% extrapolation in terms of $\beta$ to estimate the energy at lower temperatures
%% does not really resolve the problem we face.
%
%%One can overcome this problem by extrapolation in terms of 

In Appendix \ref{sec:RLM}, we present some results for the case without the sign problem,
where we find that the real Langevin simulation works
as we discussed in Section \ref{sec:review}.
%as we discussed at the beginning of Section \ref{sec:review}.

\subsection{Extrapolation with respect to the interaction term}\label{sec:t-extrapolation}

In this section, we attempt to extract the results for
larger $\beta$ (lower temperature)
by introducing an additional parameter
%%to search for a region where the CLM works and by
and making an extrapolation.\footnote{This may be compared with
%%is analogous to
the
%%ere are some approaches to overcome the sign problem,
$g$-extrapolation \cite{PhysRevLett.72.613},
% which has been used in nuclear physics,
% The idea is simple:
where one deforms the system 
by multiplying a real parameter $g$
to all the interaction terms in the Hamiltonian \eqref{Hamiltonian}
with $V_\alpha > 0$
so that the sign problem disappears for $g<0$.
Then the desired expectation value is obtained
by extrapolating the data for $g \le 0$ to $g=1$.
See Appendix \ref{sec:g-extrapolation} for more details.}
%%the discussions on the $g$-extrapolation.

%% In Appendix \ref{sec:g-extrapolation},
%% we present some results for the $g$-extrapolation for comparison,
%% where the data in the $g<0$ region are obtained correctly due to the absence
%% of the sign problem and $g=1$ corresponds to the target Hamiltonian.

Here we introduce a tunable parameter $t$ in
the Hamiltonian \eqref{Hamiltonian} as
%%define a $t$-deformed Hamiltonian as
\begin{align}
 \hat H_t
 =\sum_{i,j}T_{ij}\hat c_i^\dagger \hat c_j
 +\frac{t}{2}\sum_{\alpha} V_\alpha \left( \hat O_\alpha \right)^2 \ .
 \label{Ham-t-deform}
\end{align}
Since
%%One motivation for $t$-deformation is that
the observable at small $t$
%%can be calculated analytically by
is described by perturbation theory around $t=0$,
it is given by a polynomial in $t$
in that region.
%%unless $t$ becomes too large.
Hence we can make an extrapolation to $t=1$ using a polynomial as a natural
fitting ansatz.
To obtain the expectation value of the energy,
%%$original Hamiltonian $\hat H_{t=1}$the results for 
we calculate the expectation value of the original Hamiltonian $\hat H_{t=1}$
for various $t$.
%%use the observable $\mathcal{O}=\hat H_{t=1}$
%%and extrapolate the reliable data in the small $t$ region to $t=1$.
%%We calculate the expectation value of the original Hamiltonian $H_{g=1}$ at various $g$,

%% Another motivation is that it can be used in an integration method (see appendix \ref{integration-method} for detail {\color{red} (We will probably delete the integration method appendix section.)}).

\begin{figure}[H]
 \centering
   \includegraphics[scale=0.6]{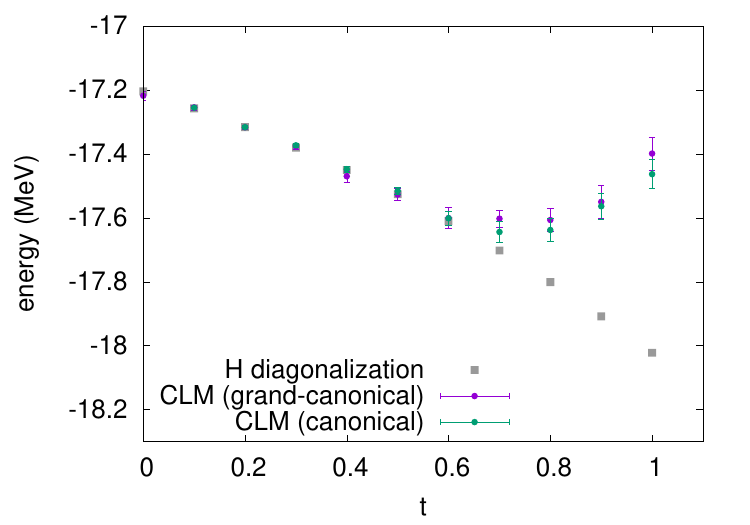}
 \includegraphics[scale=0.6]{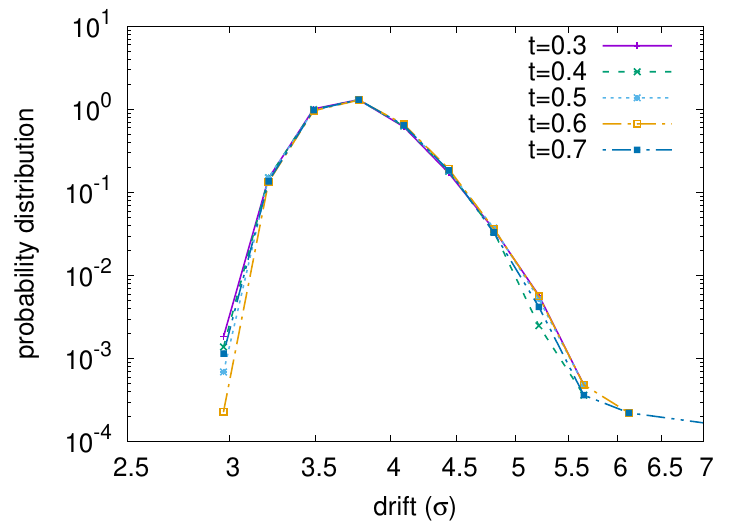}
 \includegraphics[scale=0.6]{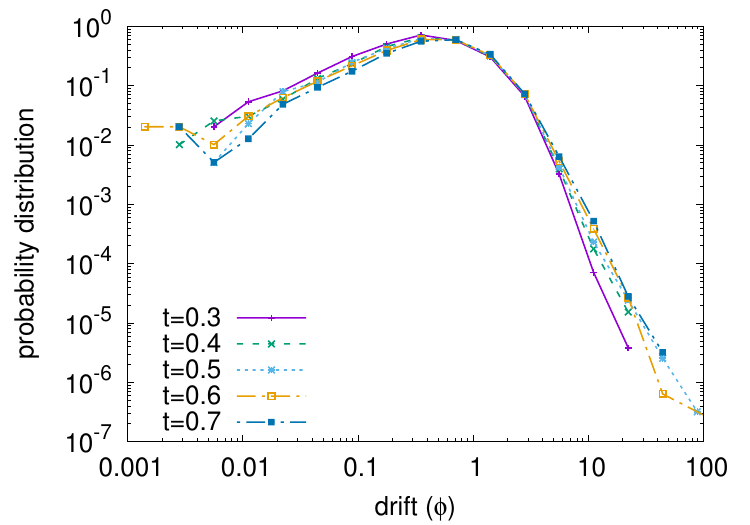}
 \includegraphics[scale=0.6]{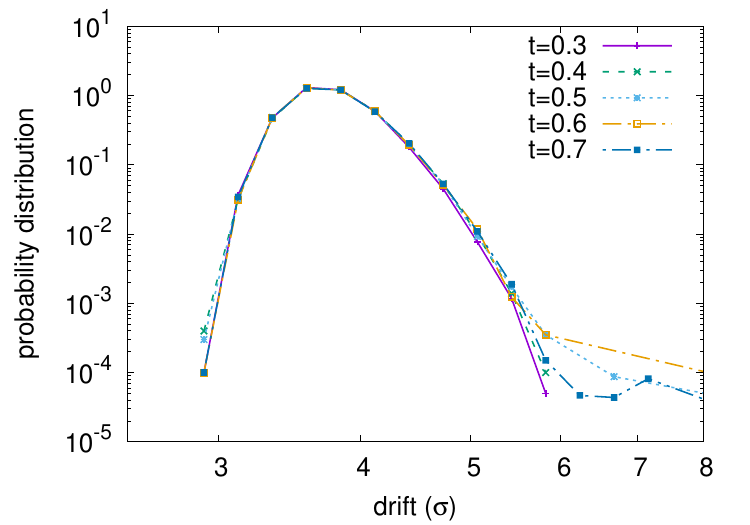}
 \caption{(Top-Left) The expectation value of the energy is plotted against the tunable
   parameter $t$
   in the canonical and grand-canonical formalisms with $N_s=20$, $N_v=2$, $\beta=1$, $n_\beta=10$.
   (Top-Right) The drift histogram for the $\sigma$ field in the canonical formalism.
   (Bottom) The drift histogram for the $\phi$ field (Left) and the $\sigma$ field (Right)
   in the grand-canonical formalism.}
 \label{fig:Comparison_Ns20_Nt10_Nv2_b1.0_tdef_Energy}
\end{figure}

%%Note we use the original Hamiltonian as an observable
%%for computation of the energy: $\mathcal{O}=\hat H_{g=1}$.

%%Our CLM simulation showed that 
%%$t$-extrapolation works in the $N_v=2$ case at $\beta=1$.
In Fig.~\ref{fig:Comparison_Ns20_Nt10_Nv2_b1.0_tdef_Energy} (Top-Left),
we plot the expectation value of the energy
against the tunable parameter $t$
for $N_v=2$ and $\beta=1$
in the canonical and grand-canonical formalisms with $n_\beta=10$.
We find good agreement with the result of the Hamiltonian diagonalization
for $t\lesssim 0.6$.
The drift histogram indicates that
the region of validity
is $t\lesssim 0.4$ for the grand-canonical formalism
and $t\lesssim 0.5$ for the canonical formalism.
%shown in Fig.~\ref{fig:NPphi_Ns20_Nt10_Nv2_b1.0_tdef_Drifts},
%%shown in Fig.~\ref{fig:Tr2_Ns20_Nt10_Nv2_b1.0_tdef_Drifts},

%%[htbp]

\begin{figure}[H]
 \centering
    \includegraphics[scale=0.6]{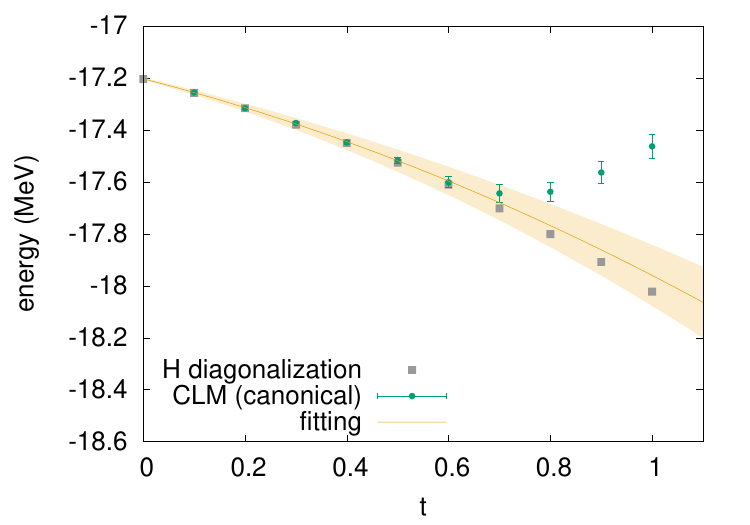}
    \includegraphics[scale=0.6]{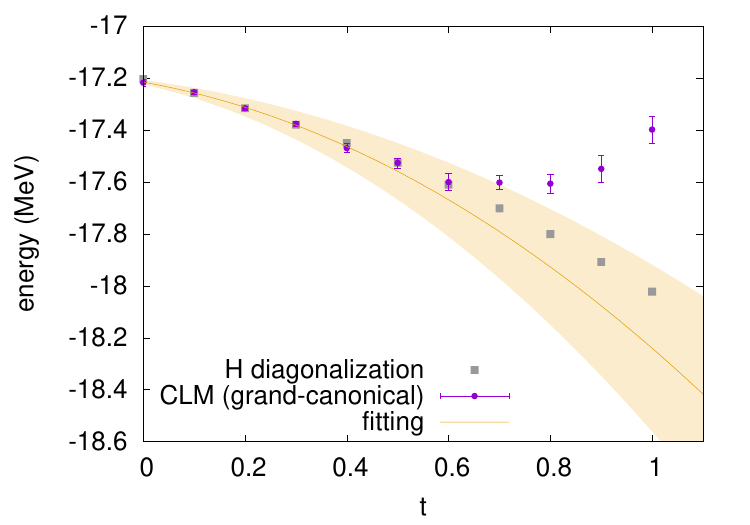}
 \caption{The expectation value of the energy is plotted against the tunable parameter $t$
   for the canonical (Left) and grand-canonical (Right) formalisms,
   with $N_s=20$, $N_v=2$, $\beta=1$ and $n_\beta=10$.
   The solid orange line represents a fit to a quadratic function,
   using the data within $0 \le t \le 0.5$ for the canonical formalism
   and the data within $0 \le t \le 0.4$ for the grand-canonical formalism.
   The shaded region represents the fitting errors. The gray points represent
   the results obtained by the Hamiltonian diagonalization, which are consistent with
   the fitting curves within the fitting errors.}
 \label{fig:Ns20_Nt10_Nv2_b1.0_tdef_Energy_fit}
\end{figure}

We fit the data in these regions to a quadratic function
in Fig.~\ref{fig:Ns20_Nt10_Nv2_b1.0_tdef_Energy_fit}.
The fitting curve obtained in this way is consistent with the result
obtained by the Hamiltonian diagonalization within the fitting errors
up to the target value $t=1$.
Thus we can obtain the expectation value of the energy at $\beta=1$.
Let us note that the difference of the energy of the ground state
and the first excited state is $\delta E \approx 1.438 ({\rm MeV})$ in the present case,
which means that $e^{-\beta \delta E} \approx 0.237\cdots$.
In order to extract the ground state energy, it would be desirable to
obtain results up to $\beta \sim 3$. The fitting ansatz in such cases would need
high orders in $t$, which makes the extrapolation less reliable.

\section{Summary}
\label{sec:summary}

We have discussed how the nuclear shell model can be investigated
without the variational ansatz
%%from first principles
by applying the CLM, which
overcomes the sign problem in quantum Monte Carlo methods.
In particular, we find 
that the grand-canonical formalism \eqref{grand-canonical-path-integral}
is convenient to simulate the case with various $N_v$
including the region $N_v \sim N_s/2$, where
the canonical formalism \eqref{fixed-Nv-path-integral} becomes too complicated to implement.
%% is not convenient
%% for large $N_v$, the validity region of the CLM turns out to be slightly larger
%% than in the grand-canonical formalism in the case studied here.

As a first step, we have performed explicit calculations
%%presented results
for ${}^{40+N_v}$Ca systems ($N_s=20$),
with a realistic potential at finite temperature.
%% We have investigated the nuclear shell model from first principles
%% using the CLM to overcome the sign problem in Monte Carlo methods.
%% In particular, we presented results for ${}^{40+N_v}$Ca systems ($N_s=20$),
%% with realistic potentials at finite temperature.
Our results agree with the results of the Hamiltonian diagonalization
at high temperature but not at low temperature.
This has been understood from the viewpoint of the validity criterion \cite{Nagata:2016vkn},
%%of the CLM,
which is a sufficient condition for the correct convergence of the CLM.
The temperature at which the agreement ceases to hold
is slightly lower than the temperature at which the drift histogram starts to have
a slow fall-off. Thus by probing the drift histogram, we can safely estimate the
validity region of the CLM calculations.

In order to obtain results at lower temperature, which is not directly accessible by
the CLM, we have proposed to introduce a tunable parameter $t$ in the interaction
terms so that the original Hamiltonian is obtained at $t=1$.
In the case we studied, the region of $t$ in which the CLM works is large enough to
%%in the grand-canonical and also the canonical formalisms.
make an extrapolation to $t=1$ by a quadratic fit.
The result obtained by this extrapolation is consistent with
the result obtained by the Hamiltonian diagonalization within the fitting errors.
The extrapolation turns out to be easier for
the canonical formalism \eqref{fixed-Nv-path-integral}
than the grand-canonical formalism \eqref{grand-canonical-path-integral}
due to a slightly larger validity region of the CLM.
%% is convenient to simulate the case with various $N_v$.
%% While the canonical formalism \eqref{fixed-Nv-path-integral} is not convenient
%% for large $N_v$, the validity region of the CLM turns out to be slightly larger
%% than in the grand-canonical formalism in the case studied here.

For
%%larger
heavier
nuclei, the computational cost increases as
%%$\sim (N_s)^p (N_v)^q n_\beta ^r$.
$\sim  (N_s)^5 (N_v)^0 n_\beta$
as we discussed at the end of section \ref{sec:app-shell-model}.
While the sign problem becomes severer, it is possible that
the validity region of the CLM is not significantly affected by the system size.
We therefore expect that the CLM can be a useful approach
to
%%large
heavy
nuclei, which is complementary to the variational method.
%% By comparisons between the results of the CLM and the Hamiltonian diagonalization,
%% we clarified that the CLM is applicable to some nuclear systems
%% and figured out to what extent the method can correctly compute 
%% observables such as the energy.
%% What is important in our results is that
%% the reliability of the CLM results can be judged 
%% by histograms of the drift term and the measured observable.
%% This is significant especially when we apply the CLM to nuclear systems 
%% the model spaces of which are too large to use the Hamiltonian diagonalization.

It is also worthwhile to mention 
a potential relation of our study to quantum computing.
Recent developments in quantum computing provide another possible route 
to addressing the rapid growth of the Hilbert space 
in nuclear shell-model calculations. 
While present-day quantum devices are still limited 
in the size and accuracy of feasible calculations, 
quantum algorithms have already been investigated for nuclear structure 
and shell-model problems \cite{Stetcu:2021cbj,Perez-Obiol:2023vod}. 
Classical methods with polynomial computational scaling, 
such as the approach investigated here, may also 
provide useful benchmarks for future quantum calculations.

\section*{Acknowledgment}

The computations were carried out on
the PC clusters in KEK Computing Research Center
and KEK Theory Center.
This research was supported by MEXT 
as ``Program for Promoting Researches on the Supercomputer Fugaku''
(Simulation for basic science: from fundamental laws of particles to creation of nuclei).
%% ``Program for promoting research on
%% the supercomputer Fugaku'', MEXT, Japan (JPMXP1020230411)
It is also supported by Joint Institute for Computational Fundamental Science (JICFuS).
Y.A.~was supported by JSPS KAKENHI Grant Number JP24K07036.

\appendix
\section{Canonical formalism}
\label{sec:appendix}
In this appendix, we discuss how to implement the canonical formalism
in the CLM. As we did for the grand-canonical formalism,
we change variables as
$\sigma_{\alpha n}=\frac{\tilde{\sigma}_{\alpha n}}{\sqrt{\iDelta\beta |V_{\alpha}|}}$
for simplicity.

We can write the effective action of the path integral
%%\eqref{fixed-Nv-path-integral} as
\eqref{def-Z-canonical} as
\begin{align}
 S=\sum_{\alpha,n}\frac{\tilde{\sigma}_{\alpha n}^{2}}{2}
 -\ln\Tr_{N_v} \hat U \  .
 \label{eff-action-canonical}
\end{align}
The complex Langevin equation is then written as 
\begin{align}
 \der{\tilde{\sigma}_{\alpha n}}{t}
 =-\pder{S}{\tilde{\sigma}_{\alpha n}}+\eta_{\alpha n}(t) \ ,
\end{align}
where $\eta_{\alpha n}$ is a Gaussian noise term that satisfies
$\langle \eta_{\alpha n}(t)\eta_{\beta m}(0) \rangle =2\delta_{\alpha\beta}\delta_{nm}\delta(t)$.

Unlike the grand-canonical formalism,
the effective action \eqref{eff-action-canonical}
involves a trace in the subspace of the Fock space with $N_v$ particles.
Therefore, in order to derive a more explicit form of the drift term,
we need to specify the value of $N_v$
and to transform $\Tr_{N_v}$
%% , defined as a trace in the Fock space with $N_v$ particles,
into an expression written in terms of matrix traces.
For instance, in the case of $N_v=2$, $\Tr_{2}\hat U$ can be rewritten as
\begin{align}
 \Tr_{2} \hat U
 =\frac{1}{2}\left[
 (\tr U)^2 - \tr U^2
 \right] \  ,
\end{align}
where $U$ is the one-body matrix defined by $U_{ij}=(\exp[-\beta h^{\rm (tot)}])_{ij}$,
and the drift term becomes
\begin{align}
 \pder{S}{\tilde{\sigma}_{\alpha n}}
 &=\tilde{\sigma}_{\alpha n}
 -\frac{2}{(\tr U)^2-\tr U^2}
 \tr \left[
 (\tr U)\pder{U}{\tilde{\sigma}_{\alpha n}}
 -\left( U\pder{U}{\tilde{\sigma}_{\alpha n}} \right)
 \right] \ .
 \label{eq:drift-canonical}
\end{align}
The derivative of $U$ with respect to $\tilde{\sigma}_{\alpha n}$
is written explicitly as
\begin{align}
 \pder{U}{\tilde{\sigma}_{\alpha n}}
 % &=-\frac{s_\alpha \iDelta\beta V_\alpha}{\sqrt{\iDelta\beta |V_\alpha|}}
 % u_{n_\beta}\cdots u_{n+1} 
 % \int_0^1 d\tau\, u_{n}^\tau O_\alpha u_{n}^{-\tau}
 % u_{n}u_{n-1}\cdots u_1
 % \nonumber \\
 &=s_\alpha^\ast \sqrt{\iDelta\beta |V_\alpha|}
 e^{-\iDelta\beta h_{n_\beta}}\cdots e^{-\iDelta\beta h_{n+1}}
 % \int_0^1 d\tau\, (e^{-\iDelta\beta h_{n}\tau}) O_\alpha 
 \left(
 \frac{e^{-\iDelta\beta \operatorname{adj}h_{n}}-1}{-\iDelta\beta\operatorname{adj}h_{n}} O_\alpha 
 \right)
 e^{-\iDelta\beta h_{n}}%e^{-\iDelta\beta h_{n-1}}
 \cdots e^{-\iDelta\beta h_1} \ ,
\end{align}
where
$\hat h_n=\sum_{i,j}(h_n)_{ij}\hat c_i^\dagger\hat c_j$ and
$\hat O_\alpha=\sum_{i,j}(O_\alpha)_{ij}\hat c_i^\dagger\hat c_j$.
We have defined an operation $(\operatorname{adj}A) \, B \equiv [A,B]$.
The appearance of $s_\alpha^\ast$ is due to the use of the relation
$s_\alpha V_\alpha  = - s_\alpha^\ast |V_\alpha|$.
By diagonalizing $h_n$ as $(P_n^{-1} h_n P_n)_{ij}=\lambda_{n,i}\delta_{ij}$,
we can rewrite the factor with $O_\alpha$ into an explicit form
\begin{align}
 \left(
 \frac{e^{-\iDelta\beta \operatorname{adj}h_{n}}-1}{-\iDelta\beta\operatorname{adj}h_{n}} O_\alpha 
 \right) _{ij}  =
 %%\sum_{i',j',k}
 \sum_{i',j'}
 (P_n)_{ii'}\frac{e^{-\iDelta\beta (\lambda_{n,i'}-\lambda_{n,j'})}-1}{-\iDelta\beta (\lambda_{n,i'}-\lambda_{n,j'})}
 %% (P_n^{-1} O_\alpha P_n)_{j'k} (P_n^{-1})_{kj}
  (P_n^{-1} O_\alpha P_n)_{i'j'} (P_n^{-1})_{j'j}
 \  .
\end{align}

\section{Some results for the case without the sign problem}
\label{sec:RLM}

\begin{figure}[thbp]
 \centering
 \includegraphics[scale=0.8]{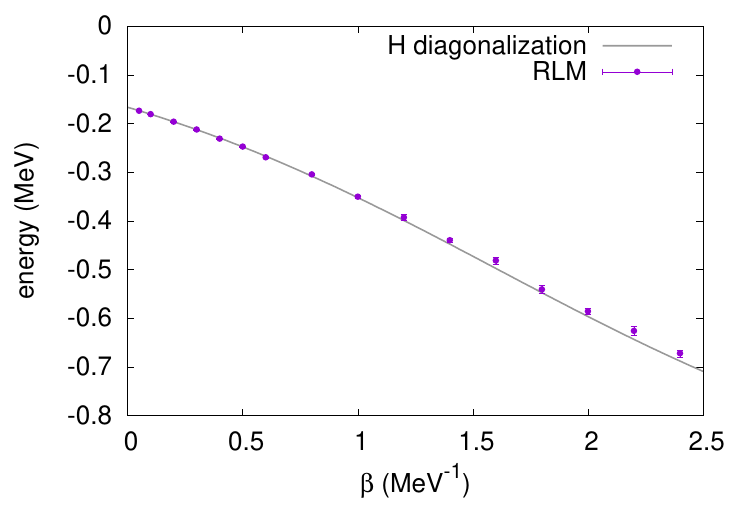}
 \caption{The expectation value of the energy is plotted against the inverse 
   temperature $\beta$ in the grand-canonical formalism with $N_s=4$, $N_v=2$, $n_\beta=10$,
   normalizing the ground state energy to be $-1$ MeV.
  The purple points with error bars represent the results of real Langevin simulations,
  whereas the gray curve represents the result obtained by the Hamiltonian diagonalization.}
 \label{fig:Cano_Ns4_Nt10_Nv2_Energy}
\end{figure}

% It would, I believe, be instructive and a nice to demonstrate the
% real Langevin equation in action on a good-sign problem. Thus, showing
% that the overall approach is justified and makes sense.
In this appendix, we show some results for the case without the sign problem
to confirm that the real Langevin simulation works in that case.
We use the same code as the one used for the complex Langevin simulations in this paper
since Eq.~\eqref{CLangevin-eq} reduces to Eq.~\eqref{RLangevin-eq}
for the real initial configuration $z(0)=x(0)$.
%% In that case, configurations are generated by Eq.~\eqref{RLangevin-eq},
%% which is equivalent to the complex Langevin dynamics by 
%% with a real initial configuration of $x$,
%% if the action $S(x)$ is real for real $x$.
%% Therefore, we can use the same code for a real Langevin simulation.

Let us consider the case with $N_s=4$ in a $p_{3/2}$ orbit
%%%shell
and $N_v=2$ as an example.
The two-particle model space consists of a $J=0$ singlet and
a $J=2$ quintuplet.
For a rotationally invariant shell-model Hamiltonian in this space,
the two-particle energies are given by $E(J)=2e_1+V_J$ with $J=0,2$,
where $e_1$ is the one-body energy and $V_J$ is the
strength of the two-body interaction for the two particles coupled with
the total angular momentum $J$.
%%In the simple case considered here, where two identical particles occupy
%%a single $j=3/2$ orbit,
Since only the relative energy $E(0)-E(2)=V_0-V_2$ is relevant,
we may set $e_1=0$ and choose $V_2=0$ as an energy reference.
The additional choice $V_0=-1\mbox{~MeV}$ merely fixes the energy scale and corresponds
to a simple attractive interaction in the $J=0$ channel.
In particular, there is no sign problem in this case.
%%the ground state energy is normalized becomes $-1$ MeV.

In Fig.~\ref{fig:Cano_Ns4_Nt10_Nv2_Energy},
we show the energy obtained by real Langevin simulations,
%%for $N_v=2$,
where we observe good agreement with the exact values at high temperature.
Slight deviation at lower temperature can be attributed to
% finiteness of $N_\beta$
the finite discretization effects of $\beta$ and the Langevin time.
This figure should be compared with Figs.~\ref{fig:NPphi_Ns20_Nt10_Nv2_Energy} (Top)
and \ref{fig:NPphi_Ns20_Nt10_Nv10_Energy} (Top),
which show the results obtained by complex Langevin simulations
for $N_s=20$, $N_v=2,10$ and $n_\beta=10$.

\begin{figure}[t]
 \centering
 \includegraphics[scale=0.8]{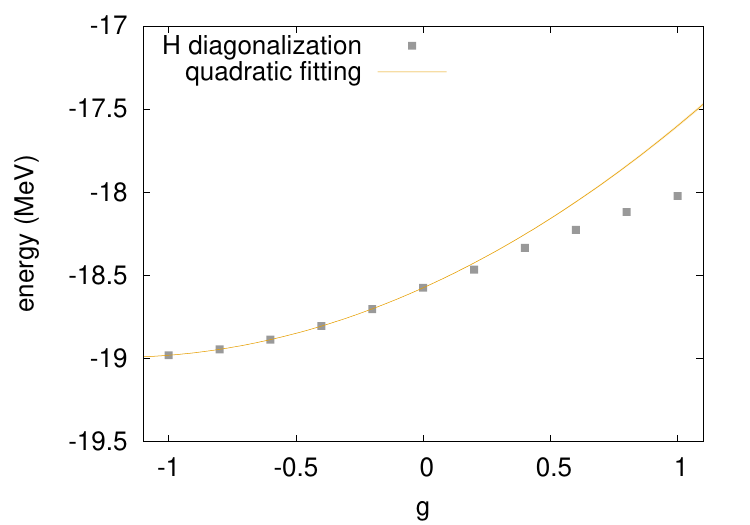}
 \caption{The expectation value of the energy obtained by the Hamiltonian diagonalization
   with $N_s=20$, $N_v=2$, $\beta=1$ is plotted against the tunable parameter $g$.
   The data within the $g \le 0 $ region
   are fitted to a quadratic function.}
 %%(Left) and a cubic function (Right).}
 \label{fig:Ns20_Nt10_Nv2_b1.0_gdef}
\end{figure}

\section{Comparison with the $g$-extrapolation}
\label{sec:g-extrapolation}
% If I understand properly, the scaling factor t is applied to the
% total two-body interaction, thus when t = 0, there are just the
% single-particle energies contribute. This states to analogous to the
% approach in Ref. [10], although I believe it is somewhat different.
% Indeed, it would be interesting to something similar to Ref. [10] by
% scaling only the components of the Hamiltonian with a bad sign and
% then extrapolating just the bad sign part. Would this work just as
% well, or worse? Here, one seems to be extrapolating the full two-body
% interaction, but possibly part of it, the good-sign part, could be
% done.

In this appendix,
we discuss the $g$-extrapolation \cite{PhysRevLett.72.613}
as an alternative way to perform extrapolations
from the region in which correct results are available.
%% In particular, we compare it with
%% the extrapolation by the real parameter $t$
%% used in section \ref{sec:t-extrapolation}.

Here we pay attention to the fact that
the terms with $V_\alpha>0$ in the Hamiltonian \eqref{Hamiltonian}
are the ``bad'' interactions,
which produce imaginary factors in the partition function
through $s_\alpha=i$ in \eqref{def-h-hat}.
This motivates us to multiply a tunable parameter $g$ only to the interaction terms
with $V_\alpha > 0$ in \eqref{Hamiltonian}
so that the $g$-deformed Hamiltonian reads
\begin{equation}
 \hat H_g
 =f(g)\left(
 \sum_{i,j=1}^{N_s}T_{ij}\hat c_i^\dagger \hat c_j
 +\frac{1}{2}\sum_{\alpha, V_\alpha<0} V_\alpha \left( \hat O_\alpha \right)^2
 \right)
 +\frac{g}{2}\sum_{\alpha, V_\alpha>0} V_\alpha \left( \hat O_\alpha \right)^2 
 \  ,
 \label{Ham-g-deform}
\end{equation}
which is in contrast to the $t$-deformed Hamiltonian \eqref{Ham-t-deform},
%%used in section \ref{sec:t-extrapolation},
where all the interaction terms are commonly multiplied by $t$.
In Eq.~\eqref{Ham-g-deform},
we have introduced an arbitrary function $f(g)$ satisfying $f(g)>0$ for $g<0$ and $f(1)=1$,
which is set to $f(g)=1$ in what follows for simplicity.
%%so that $g=1$ retrieves the original Hamiltonian.
%% Let us set $f(g)$ to unity in what follows for simplicity.
%% does not suffer from the sign problem for $g\leq 0$.
Note that the $g$-deformed Hamiltonian \eqref{Ham-g-deform}
does not suffer from the sign problem for $g\leq 0$.
To obtain the expectation value of the energy,
we calculate the expectation value of the original Hamiltonian $H_{g=1}$ for various $g$.
%%observable $\mathcal{O}=\hat H_{g=1}$
%%and extrapolate the data for $g\leq 0$ to $g=1$.
%% to calculate the energy for the $g$-deformed Hamiltonian.
%%use the original Hamiltonian as an observable
%%for computation of the energy: 

%%[htbp]

In Fig.~\ref{fig:Ns20_Nt10_Nv2_b1.0_gdef}, we plot
the expectation value of the energy obtained by the Hamiltonian diagonalization
against $g$ for the same setting as in Fig.~\ref{fig:Ns20_Nt10_Nv2_b1.0_tdef_Energy_fit}.
We fit the results for $g\le 0$ to a quadratic function,
%%up to the second (Left) and third (Right) order,
where the error bands represent the fitting errors.
%% The errors of the fitting curves are
%% estimated by deviation of fitting functions 
%% obtained from the points with $-1\leq g\leq 0$ (Left)
%% and from those with $-1\leq g\leq 0.2$ (Right), respectively.
We find that the fit does not reproduce the correct value at $g=1$.
%% both the quadratic fit (Left) the cubic fit (Right) 
%%in Fig.~\ref{fig:Ns20_Nt10_Nv2_b1.0_gdef}
%% does not give the correct value at $g=1$,
%% while the cubic fit (Right) gives a value consistent with the
%% exact result at $g=1$ albeit with a large fitting error.
This has something to do with the existence of an inflection point
near $g=0$, which does not exist in the
$t$-extrapolation in Fig.~\ref{fig:Ns20_Nt10_Nv2_b1.0_tdef_Energy_fit}.
%(See Fig.~\ref{fig:Ns20_Nt10_Nv2_b1.0_tdef_Energy_fit}.)
While this may not be a generic feature of the
$g$-extrapolation, we consider that the $t$-extrapolation has a certain
advantage at least in the case at hand.

%% Since in practical cases such as computations by the SMMC method or the CLM,
%% the fitting errors should become larger because of statistical errors,
%% we chose the extrapolation method by the parameter $t$ 
%% in section \ref{sec:t-extrapolation}.

\bibliographystyle{JHEP}
\bibliography{CLSM}

\end{document}